\documentclass[twocolumn,showpacs,preprintnumbers,amsmath,amssymb]{revtex4}

\usepackage{graphicx}
\usepackage{dcolumn}
\usepackage{bm}

\newcommand\ignore[1]{}
\def\one{{\,\hbox{1\kern-.8mm l}}}

\def\Tr{{\rm Tr\, }}

\newcommand{\Cset}{{\,\,{{{^{_{\pmb{\mid}}}}\kern-.45em{\mathrm C}}}}}

\newcommand{\be}{\begin{equation}}
\newcommand{\ee}{\end{equation}}
\newcommand{\bea}{\begin{eqnarray}}
\newcommand{\eea}{\end{eqnarray}}

\def\a{\alpha}\def\b{\beta}

\def\d{\partial}

\begin{document}

\title{Towards a Realization of the Condensed-Matter/Gravity Correspondence in String Theory via Consistent Abelian Truncation}

\author{Asadig Mohammed$^{1}$}\email{asadig@gmail.com}
\author{Jeff Murugan$^{1}$}\email{jeff@nassp.uct.ac.za}
\author{Horatiu Nastase$^{2}$}\email{nastase@ift.unesp.br}
\affiliation{${}^{1}$The Laboratory for Quantum Gravity \& Strings, Department of Mathematics and Applied Mathematics, University of Cape Town,Private Bag, Rondebosch 7700, South Africa}
\affiliation{${}^{2}$Instituto de F\'{i}sica Te\'{o}rica, UNESP-Universidade Estadual Paulista, Rua Dr. Bento T. Ferraz 271, Bl. II, S\~ao Paulo 01140-070, SP, Brazil}
\date{\today}

\begin{abstract}
We present an embedding of the 3-dimensional relativistic Landau-Ginzburg model for condensed matter systems in 
an $\mathcal{N}=6$, $U(N)\times U(N)$ Chern-Simons-matter theory (the ABJM model) by consistently truncating the latter to an abelian effective field theory encoding the collective dynamics of ${\cal O}(N)$ of the ${\cal O}(N^2)$ modes. In fact, depending on the VEV on one of the ABJM scalars, a mass deformation parameter $\mu$ and the Chern-Simons level number $k$, our abelianization prescription allows us to interpolate between the abelian Higgs model with its usual multi-vortex solutions and a $\phi^4$ theory. We sketch a simple condensed matter model that reproduces all the salient features of the abelianization. In this context, the abelianization can be interpreted as giving a dimensional reduction from four dimensions.
\end{abstract}

\pacs{11.25.Tq, 11.15.Yc, 74.40.Kb}

\maketitle

\section{Introduction}
The gauge/gravity duality, as manifest in the AdS/CFT correspondence \cite{Maldacena} 
has evolved from its humble origins in string theory into one of the most powerful tools in the arsenal of physicists studying non-perturbative and strong-coupling phenomena in quantum field theories today. 
The original and, arguably, most studied of these "applied string theory" phenomena was the physics of QCD. In particular, a lot of recent work was 
focused on the quark-gluon plasma observed in heavy ion colliders such as the RHIC collider and the ALICE experment at CERN 
(see, for example, \cite{Gubser} for a recent review). More recently though, the ideas of holography have found a new, lower-dimensional hunting ground in {\it condensed matter physics}. The pp-wave or BMN limit of AdS/CFT selects operators with large charge \cite{Berenstein:2002jq} in $\mathcal{N}=4$ super Yang-Mills theory. The physics of such large operators is, in a very concrete sense, isomorphic to the physics of certain spin chains \cite{Minahan:2002ve}; a realization that has led, not only to an enormous development in our understanding of integrability in string theory but also served as a forerunner to many of the developments in what has now become known as the ``AdS/CMT" correspondence. 

In these cases however, the duality was {\it heuristically motivated} by a relation between some system of branes and a gravitational background, in some decoupling limit. Later, it was realized that if physics in AdS is always holographic, we can consider simple theories in AdS, that should be related to some strongly coupled conformal field theory on the boundary. Thus naturally AdS/CFT came to be applied to condensed matter systems, where one encounters strongly coupled conformal field theories that cannot be dealt with in other ways (see for instance the reviews 
\cite{Hartnoll:2009sz,Herzog:2009xv} for an introduction and relevant references). In all these cases, however, the argument is mostly one of universality, that a variety of (large $N$) theories have some small set of (abelian) operators dual to some finite and small number of fields in AdS, usually a $U(1)$ gauge field, a scalar and maybe some spinors, representing a (sometimes consistent) truncation of some AdS/CFT pair. In other words, either 
one truncates the number of operators of the system, in which case it is not entirely clear {\it (a)} why one should focus on a subset, or how one understands from the point of view of a condensed matter system the focus on the few operators of the large $N$ system; or {\it (b)} one thinks of an abelian condensed matter {\it analog} of the large $N$ theory, in which case it is not clear why we can use just a gravity dual, as opposed to a full string 
theory. In either case, we find the argument less than persuasive.

In this note we will take some steps towards a better understanding of AdS/CMT, by proposing a modification of the above set-up. We consider instead a {\it consistent truncation} of the 3-dimensional ABJM theory (which has a known gravity dual), a truncation that corresponds to the collective dynamics 
of ${\cal O}(N)$ fields out of the ${\cal O}(N^2)$ of ABJM, and gives an effective theory that is easily identified as the relativistic Landau-Ginzburg model. We also sketch a simple CMT model that has the same qualitative features as the ABJM abelianization, allowing us to understand better in what sense can we use ABJM for condensed matter systems. Here we present only the main ideas, leaving the technical details to a longer paper \cite{soon}.

As a point of clarity, we note that the idea of a consistent truncation in string theory is not a new one, having featured before in two primary contexts. On the gravity side of the AdS/CFT correspondence, when one is interested in a classical limit only, a consistent truncation means that we can safely drop the ``nonzero modes", as 
these will only appear in quantum loops. In supergravity compactifications however, a consistent truncation for a 
dimensional reduction means that we can drop all the nonzero (KK) modes from the low energy quantum theory as well, provided that the coupling to the nonzero modes can be made arbitrarily small, or that the masses of the nonzero modes are much larger than the mass parameters of the low energy theory. Hence quantum loops of these
nonzero modes may be ignored. We will argue below that it is this latter case that arises arises here. 
To understand the collective dynamics that is crucial to our argument, consider a large number, $N$, 
of branes in some background. A classical solution that is obtained by turning on fields in all the $N$ branes 
will curve the background space nontrivially corresponding, via AdS/CFT, to some finite deformation of the 
dual theory. On the other hand, just turning on fields on a single brane will produce a negligible deformation 
that will not deform the background space nor the dual. In our case it is the former situation that arises so that in this sense, the collective dynamics of ${\cal O}(N)$ fields really is different from the dynamics of a single 
field, and it is this that allows for a dual gravitational interpretation.

\section{ABJM and its massive deformation}

The ABJM model \cite{Aharony:2008ug} 
is an ${\cal N}=6$ supersymmetric $U(N)\times U(N)$ Chern-Simons gauge theory at level $(k,-k)$, with bifundamental scalars 
$C^I$ and fermions $\psi_I$, $I=1,...,4$ in the fundamental of the $SU(4)_R$ symmetry group and gauge fields for the two groups 
$A_\mu$ and $\hat A_\mu$. Its action is given by
\begin{widetext}
\bea
S&=&\int d^3x \Bigg( \frac{k}{4\pi}\epsilon^{\mu\nu\lambda}\Tr\Big(A_{\mu}\partial_{\nu}A_{\lambda}+\frac{2i}{3}A_{\mu}A_{\nu}A_{\lambda} -\hat{A}_{\mu}\partial_{\nu}\hat{A}_{\lambda}
      -\frac{2i}{3}\hat{A}_{\mu}\hat{A}_{\nu}\hat{A}_{\lambda}\Big)
-\Tr D_{\mu}C_{I}^{\dagger}D^{\mu}C^{I}
      -i\Tr \psi^{I\dagger}\gamma^{\mu}D_{\mu}\psi_{I}\nonumber\\
      &+&\frac{4\pi^2}{3k^2}\Tr \Big(C^{I}C_{I}^{\dagger}C^{J}C_{J}^{\dagger}C^{K}C_{K}^{\dagger}
+C_{I}^{\dagger}C^{I}C_{J}^{\dagger}C^{J}C_{K}^{\dagger}C^{K}
      +4C^{I}C_{J}^{\dagger}C^{K}C_{I}^{\dagger}C^{J}C_{K}^{\dagger}
      -6C^{I}C_{J}^{\dagger}C^{J}C_{I}^{\dagger}
      C^{K}C_{K}^{\dagger}C^{K}\Big)\label{abjmaction}\\
&+&\frac{2\pi i}{k}\Tr\Big(C_{I}^{\dagger}C^{I}\psi^{J\dagger}\psi_{J}-\psi^{J\dagger}C^{I}C_{I}^{\dagger}\psi_{J}
      -2C_{I}^{\dagger}C^{J}\psi^{I\dagger}\psi_{J}
      +2\psi^{J\dagger}C^{I}C_{J}^{\dagger}\psi_{J}
+\epsilon^{IJKL} C_{I}^{\dagger}\psi_{J}C_{K}^{\dagger}\psi_{L}-\epsilon_{IJKL}C^{I}\psi^{J\dagger}C^{K}\psi^{L\dagger}\Big)\Bigg),\nonumber
\eea
\end{widetext}
where the gauge-covariant derivative is 
\be
      D_{\mu} C^{I}=\partial_{\mu}C^{I}+iA_{\mu}C^{I}-iC_{I}\hat{A}_{\mu}.
\ee
The action has an $SU(4)\times U(1)$ R-symmetry associated with the ${\cal N}=6$ supersymmetries. It admits a 
maximally supersymmetric ({\em i.e.}, preserving all ${\cal N}=6$) massive deformation with mass parameter $\mu$ \cite{Gomis:2008vc,Terashima:2008sy}, 
which breaks the R-symmetry down to $SU(2) \times SU(2)\times U(1)_{A}\times U(1)_{B}\times \mathbb{Z}_{2}$ by splitting the scalars as
\be
     C^{I}=(Q^{\alpha},R^{\alpha}); \qquad \alpha=1,2
\ee
The mass deformation changes the potential to 
\be
     V=\Tr\left(|M^{\alpha}|^2+|N^{\alpha}|^2\right),\label{masspot}
\ee
where
\begin{widetext}
 \bea
 M^{\alpha}&=& \mu Q^{\alpha}+\frac{2\pi}{k}\Big(2Q^{[\alpha}Q^{\dagger}_{\beta}Q^{\beta]}+R^{\beta}R^{\dagger}_{\beta}Q^{\alpha}-Q^{\alpha}R^{\dagger}_{\beta}R^{\beta}
   +2Q^{\beta}R^{\dagger}_{\beta}R^{\alpha}-2R^{\alpha}R^{\dagger}_{\beta}Q^{\beta}\Big),\nonumber\\
   N^{\alpha} &=& -\mu R^{\alpha}+\frac{2\pi}{k}\Big(2R^{[\alpha}R^{\dagger}_{\beta}R^{\beta]}+Q^{\beta}Q^{\dagger}_{\beta}R^{\alpha}-R^{\alpha}Q^{\dagger}_{\beta}Q^{\beta}
   +2R^{\beta}Q^{\dagger}_{\beta}Q^{\alpha}-2Q^{\alpha}Q^{\dagger}_{\beta}R^{\beta}\Big).\cr
&&\label{mandn}
 \eea
\end{widetext} 
This mass-deformed theory (mABJM) has ground states of the {\it fuzzy sphere} type given by 
\cite{Gomis:2008vc,Terashima:2008sy}
\be
R^\a=c G^\a;\;\;\; Q^\a=0\;\;\;{\rm and}\;\;\;
Q^\dagger_\a=c G^\a;\;\;\; R^\a=0
\ee
where $c\equiv\sqrt{\frac{\mu k}{2\pi}}$
and the matrices $G^\a$, $\a=1,2$, bifundamental under $U(N)\times U(N)$, satisfy (with no summation on repeated indices)
\be
G^\a=G^\a G^\dagger_\b G^\b-G^\b G^\dagger_\b G^\a
\ee
This ground state corresponds to a fuzzy 2-sphere \cite{Nastase:2009ny,Nastase:2010uy}.
An explicit solution for $G^\a$ is given by 
\bea\label{BPSmatrices}
( G^1)_{m,n }    = \sqrt { m- 1 } ~\delta_{m,n},\quad
( G^2)_{m,n} = \sqrt { ( N-m ) } ~\delta_{ m+1 , n } \nonumber\\
\\
 (G_1^{\dagger} )_{m,n} = \sqrt { m-1} ~\delta_{m,n}, \quad
( G_2^{\dagger} )_{m,n} = \sqrt { (N-n ) } ~\delta_{ n+1 , m }.\nonumber
\eea
We will now use these so-called GRVV matrices to posit an ansatz that effectively abelianizes the ABJM model while retaining the large $N$ limit.

\section{Consistent Abelian Truncation and Condensed Matter Model}

Consider then the following ansatz for the Chern-Simons fields and the scalar matter in the supermultiplet:
\bea
A_{\mu}&=&a^{(2)}_{\mu}G^{1}G_{1}^{\dagger}+a^{(1)}_{\mu}G^{2}G_{2}^{\dagger}\nonumber\\
\hat{A}_{\mu}&=&a^{(2)}_{\mu}G_{1}^{\dagger}G^{1}+a^{(1)}_{\mu}G_{2}^{\dagger}G^{2}\nonumber\\
Q^{\alpha}&=&\phi_{\alpha}G^{\alpha}\nonumber\\
R^{\alpha}&=&\chi_{\alpha}G^{\alpha},
\eea
where, again, there is no summation over the repeated $\a$;
$a_{\mu}^{(1)}$ and $a_{\mu}^{(2)}$ are real-valued vector fields and $\phi_{\alpha}$, $\chi_{\alpha}$  are complex-valued scalar fields. 
This provides a consistent truncation of the ABJM action to 
\bea 
S&=&-\frac{N(N-1)}{2}\int d^{3}x\Bigg[\frac{k}{4\pi}\epsilon^{\mu \nu \lambda}\big(a^{(2)}_{\mu}f^{(1)}_{\nu \lambda}+a^{(1)}_{\mu}f^{(2)}_{\nu \lambda}\big)\nonumber\\
&+&|D_{\mu}\phi_{i}|^{2}+|D_{\mu}\chi_{i}|^{2}+U(|\phi_{i}|,|\chi_{i}|)\Bigg]\label{abelianmaster}
\eea
where $U\equiv 2V/N(N-1)$ is a rescaling of the potential 
\bea
V&=&\frac{2\pi^{2}}{k^2}N(N-1)\Big[(|\phi_{1}|^{2}+|\chi_{1}|^{2})\big(|\chi_{2}|^{2}-|\phi_{2}|^{2}-c^{2}\big)^{2}\nonumber\\
&+&(|\phi_{2}|^{2}+|\chi_{2}|^{2})\big(|\chi_{1}|^{2}-|\phi_{1}|^{2}-c^{2}\big)^{2}\\
&+&4|\phi_{1}|^{2}|\phi_{2}|^{2}(|\chi_{1}|^{2}\nonumber\\
&+&|\chi_{2}|^{2})+4|\chi_{1}|^{2}|\chi_{2}|^{2}(|\phi_{1}|^{2}+|\phi_{2}|^{2})\Big],\nonumber
\label{abelianpot}
\eea
and where the, now abelian, gauge covariant derivatives are $D_{\mu}\phi_{i}=(\partial_{\mu}-ia_{\mu}^{(i)})\phi_{i}$ and $D_{\mu}\chi_{i}=(\partial_{\mu}-ia_{\mu}^{(i)})\chi_{i}$. Different choices of scalars turned on lead to different consistent truncations we 
collect below: 
\begin{itemize}
  \item \underline{$\chi_2=\phi_2=0$:} This leads to a model with two massive complex scalars with no self-interactions. This is essentially trivial and will not merit further attention.
   \item \underline{$\chi_1=\phi_2=0$:} After a minor re-labeling of $\chi_2\rightarrow\phi_2$ this choice produces
\bea
S&=&-\frac{N(N-1)}{2}\int d^{3}x\Bigg[\frac{k}{4\pi}\epsilon^{\mu \nu \lambda}\big(a^{(2)}_{\mu}f^{(1)}_{\nu \lambda}+a^{(1)}_{\mu}f^{(2)}_{\nu \lambda}\big)\nonumber\\
&+&|D_{\mu}\phi_{i}|^{2} +U(|\phi_{i}|)\Bigg],\\
V&=&\frac{2\pi^2 N(N-1)}{k^2}\Bigg[|\phi_1|^2(|\phi_2|^2-c^2)^2
+|\phi_2|^2(|\phi_1|^2+c^2)^2\Bigg]\nonumber
\label{newhiggs}
\eea
a model which has vortex solutions with $\phi_1=|\phi_1|e^{iN_1\alpha}$ and $\phi_2=|\phi_2|e^{iN_2\alpha}$, where $\a$ is the polar angle on the plane, and $|\phi_{1,2}|$ go to zero at $r=0$ and $r=\infty$. 

  \item \underline{$\phi_1=\phi_2=0$:} If we also set $\chi_1=b=$constant, and solve for the (now auxiliary) gauge field $a_\mu^{(1)}$ we obtain the action 
\be
S=-\frac{N(N-1)}{2}\int d^3x\Big[\frac{k^2}{8\pi^2|b|^2}(f_{\mu\nu}^{(2)})^2+|D_\mu \chi_2|^2+V\Big]\label{abhiggs}
\ee
where the potential is
\bea
V &=&\frac{2\pi^2}{k^2}N(N-1)\left[|b|^2|\chi_{2}|^4+|\chi_{2}|^2((|b|^2-c^2)^2-2c^2|b|^2)\right.\cr
&&\left.+c^4|b|^2\right].\label{potLG}
\eea
This is the most interesting case.
Indeed, we see that for $||b|^2-c^2|>\sqrt{2}|c||b|$, we obtain a regular $\phi^4$ phase, whereas for $||b|^2-c^2|<\sqrt{2}|c||b|$ we obtain the abelian-Higgs
phase, i.e. this is a relativistic Landau-Ginzburg theory, with $(|b|^2-c^2)^2\sim g$ and $2c|b|^2\sim g_c$. However, in order to have a consistent
truncation, in the above action we need to also satisfy the ``equation of motion" for the constant $|b|^2$. This is indeed the case for BPS solutions of the abelian-Higgs action. 
\end{itemize}

When canonically normalizing all the fields, the quartic coupling for the {\it canonical} scalar $\tilde\chi_2$
from (\ref{potLG}) becomes $g^2$, with $g=2\pi |b|/(Nk)$, 
and the coefficient of the middle term in (\ref{potLG}) becomes $N^2g^2/2$. Then, for $|b|\sim c$, with $k\sim 1$ and 
$N$ large, $g^2\sim \mu/(N^2k)\sim \mu/N^2\ll \mu$, and generically the mass of $|\phi|$ is $\sim \mu$. But we can 
tune the system to be near zero mass, so that (for $m^2\neq0$) we have $|m^2|\ll \mu^2$. Generic 
modes of the ABJM model (the ``nonzero modes" dropped in our consistent truncation) have mass $\mu$, as easily checked
in (\ref{abjmaction}), (\ref{mandn}). 
Therefore we can drop the nonzero modes in the reduced low energy theory {\it even at the quantum level}, as advertised in the introduction, and consistently truncate to (\ref{abhiggs}). 
Note also that substituting the reduction ansatz into the ABJM action, we find that the (sextic) 
potential gives a term with bilinear coupling to the 
nonzero modes $\delta\phi$ of the type $(\tilde \chi_2)^4(\delta\phi)^2/(k^2N^2)\propto 1/N^2$, the 2-fermi-2-scalar
term gives a bilinear coupling $(\tilde\chi_2)^2\bar{\delta\psi} \delta\psi/(kN)\propto 1/N$, and mass deformation 
quartic in the scalars gives a bilinear coupling $(\tilde\chi_2)^2(\delta\phi)^2\mu/(kN)\propto \mu/N$, which is 
$\ll \mu$, though still $\gg \mu/N^2\sim g^2$. 

Now also note that for $|b|=c$, the potential (\ref{potLG}) has the vacuum $|\phi|=|b|=c$, which is nothing but the 
fuzzy sphere vacuum of the massive ABJM, therefore classical solutions of the reduced theory (\ref{abhiggs}) are 
some type of deformations of the fuzzy sphere. Other examples of such classical solutions will be given elsewhere 
\cite{soon}. These solutions, giving a collective dynamics of ${\cal O}(N)$ modes, 
then correspond to finite deformations of the gravity dual, 
unlike any solutions obtained by turning on a single mode. Therefore we retain the good features of the large $N$ behaviour (classical gravity dual) with this abelian reduction, as advertised in the introduction.

At this point, one might ask whether the modes in the Landau-Ginzburg action we consider are the {\it only} light ones and if not, do they couple to any others? For the extra modes in (\ref{abelianmaster}) that we dropped, the answer is simple. All canonically normalized fields couple to the LG mode with coupling $\sim \mu/(N^2k)\sim g^2$. Moreover,  all $\phi_i,\chi_i$ have an 
explicit mass term of the order $N^2c^4/k^2\sim \mu^2$ and there is also a contribution from the VEV $\chi_1=b$ of
$1/k^2[-4c^2b^2|\tilde\chi_2|^2+b^4(|\tilde \chi_2|^2+|\tilde \phi_2|^2)]$. Consequently in the region of parameter space where our LG mode $\tilde\chi_2$ can be tuned to be light all the other modes
{\it stay heavy}. For generic modes outside the action (\ref{abelianmaster}) the answer is a bit more difficult. As we have already argued, since generic mass terms are 
of the order $m^2\sim\mu^2>0$, the only thing remaining to check is if they can be (almost) cancelled by the terms coming from the Higgs VEV $\chi_1=b$. We can obtain 
the total mass term by keeping only two $C^I$'s in the potential (\ref{masspot}), and replacing the rest by $C^I=(R^1=bG^1,R^2=Q^1=Q^2=0)$. Setting this to 
zero produces a very long equation for the trace of products of two $C^I$ matrices and up to four $G^1$ matrices being zero, which we will not reproduce here. 
One solution is given by our LG light mode, $C^I=(R^2=\chi_2G^2,R^1=Q^1=Q^2=0)$, and amounts to an identity between the $G^1$ and $G^2$ matrices 
(together with the condition $|b|^2=c^2(2\pm \sqrt{3})$ for a massless LG field). The question is whether the solution is unique. While we don't know of a 
general {\it mathematical} proof of uniqueness, {\it physically} it is clear that there can't
be another solution. Indeed, this solution is related to the existence of the maximally supersymmetric fuzzy sphere vacuum characterized by $G^1,G^2$; once 
we have $G^1$ turned on, there is an instability towards turning on $G^2$ also. Consequently, the mass of $\chi_2$ can become negative, passing through zero. 
Another solution would amount to another instability towards a different vacuum with the same $G^1$ turned on. As there are no vacua connected in this way 
to the maximally supersymmetric one, this is impossible. Finally, we note that there can be other light modes in other regions of parameter space, but since all we 
need here is that at $N$ large, $k\sim 1$ and the only VEV turned on is $bG^1$, there are {\it no other light modes}.

The question at this stage is {\it how relevant are any of our effective field theories in the condensed matter context?} Following \cite{Sachdev:2011wg}, we now outline an argument to suggest that an appropriate answer is {\it very}. Beginning with the Hubbard model for spinless bosons, 
with a ground state where each site in the lattice is populated by an equal number of bosons, one can construct a discretized field $\phi_i\sim \a_i a_i+\b_i h_i^\dagger$, where $a_i^\dagger $ creates a ``particle" above the ground 
state and $h_i^\dagger$ creates a ``hole". One then obtains the relativistic Landau-Ginzburg action
\be
S=\int d^3x (-|\d_t\phi|^2+v^2|\vec{\nabla}\phi|^2+(g-g_c)|\phi|^2+u|\phi|^2)\label{lg},
\ee
in the continuum limit. For $g<g_c$ we have an abelian-Higgs system, i.e. a superconducting phase, while for $g>g_c$ we have an insulator phase. At $g=g_c$ (and temperature
$T=0$) the model 
describes a conformal field theory. The systems described by the above model have also a quantum critical phase which opens up at nonzero temperature for a $T$-dependent window around $g=g_c$. This quantum critical phase is strongly coupled and difficult to analyze using usual condensed matter methods and hence a good candidate for a holographic description.

Moreover, the Hubbard model is a drastic simplification of a real condensed matter system. The model was used to describe the quantum critical 
phase of (bosonic) ${}^{87}\!Rb$ cold atoms on an optical lattice, but the description is believed to hold more generally for the quantum critical 
phase. For instance, high $T_c$ superconductors have a ``strange metal" phase that is believed to be of the same quantum critical phase type.
In fact, a simple model for a solid with free electrons describes the qualitative 
features of the ABJM abelianization. One can consider an electron at site $i$ in the model coupled to an electron at site $j$ to form a spinless 
boson $\phi_{ij}=\bar\psi_i\psi_j$. In two spatial dimensions there are ${\cal O}(N^2)$ neighbours of maximum distance $N$ away. 
It is not unreasonable to consider that the length between the sites has a maximum value $N\geq |\vec{i}-\vec{j}|$. We can write this field 
as $\phi_{i'}^{ab}$, where $i'$ is at midpoint between $i$ and $j$, and $a,b$ correspond to $1,2,...,N$ in spatial directions $x,y$. 
If the normalized wavefunctions for $\phi_{i'}^{ab}$ give probabilities for existence of the pair $|\phi_{i'}^{ab}|^2$, and assuming rotational 
invariance so only rotationally invariant modes $\psi(a)$, thought of as eigenvalues of the matrices $\phi^{ab}$, are nonzero, 
we can consider a decaying solution $|\psi(a)|^2\propto N-a$. This corresponds to the ABJM matrix $G^2$, with $(G^2G^\dagger_2)_{mn}=(N-m)\delta_{mn}$.
Indeed, in ABJM we have the field $\chi_2 (G^2)_{mn}$, corresponding to $b_{i'}\sim \sum_a\psi(a)\phi_{i'}^{aa}$. 
The average distance in between sites is then 
\bea
\langle a\rangle=\frac{\int |\psi(a)|^2 a (2\pi a da)}{\int |\psi(a)|^2(2\pi a da)}=\frac{N}{2},
\eea
consistent with the fact that there is a large distance between sites that couple, as is known to be the case. 

Note that while the above fields are the only ones that are turned on, the system has, in principle, several more possibilities. For example, we can form more than one matrix scalar field, like the 4 $C^I$'s of ABJM, by having more electrons at each site that can couple to form spinless bosons, as well as 
matrix fermions, by having two electrons at site $i$ couple among themselves and with an electron at site $j$. We can also construct two Chern-Simons 
(topological) gauge fields by a generalization of the abelian CS case (see e.g. \cite{simon}) as follows: 
Consider two fermions at sites $i$ and $i''$ coupling 
to form $\phi_{i'}^{aa'}$ at their midpoint $i'$ and two fermions at sites $j$ and $j''$ coupling to form $\phi_{j'}^{bb'}$ at their 
midpoint $j'$. Then the field 
\be
e\vec{a}(\vec{r}_{i'})=\vec{\nabla}_{i'}\sum_{j'\neq i'}\a(\vec{r}_i-\vec{r}_j),\label{nonabcs}
\ee
where $\a(\vec{r}_i-\vec{r}_j)$ is the angle made by $\vec{r}_i-\vec{r}_j$ with a fixed axis, corresponds to a CS gauge field. The indices on the 
gauge field are the planar indices for the only variable above, $\vec{r}_{ii'}-\vec{r}_{jj'}$ (changing 
$\vec{r}_{ii'}$ by itself just gives a harmless overall translation), as well as the discrete choice for $\vec{r}_{ii'}$ to belong to $i'$ or $j'$,
giving two gauge fields $A$ and $\hat A$. The scalars $\phi_{i'}^{ab}$ act as bifundamental with respect to them. It is clear then that, qualitatively at least, the model outlined above describes all the fields of ABJM, as well as the abelianization. 

In some sense, the 3-dimensional Landau-Ginzburg model makes more sense as a dimensional reduction from four dimensions; the same is true of our abelianization picture. 
The matrix $G^1$, one of the two matrices $G^1,G^2$ that describe the fuzzy 2-sphere, is multiplied by the constant $|b|$, so in a sense we have 
a ``fuzzy circle" (limit of a fuzzy 2-sphere), becoming classical at large $N$. The physical radius of a fuzzy sphere construction was argued 
in the literature on brane polarizations to be (see e.g. \cite{Nastase:2009ny})
\be
R_{ph}^2=\frac{2}{N}\Tr[X^IX_I^\dagger]=\frac{2}{N}\Tr[C^IC^\dagger_I]4\pi^2l_P^3
\ee
where $l_P^3=l_s^2R_{11}$. Assuming the same formula holds for the ``fuzzy circle" case, and that 
like in the pure fuzzy sphere case, the 11th direction has radius $R_{11}=R_{ph}/k$, we obtain 
\be
R_{ph}=(N-1)l_s^2\frac{4\pi^2 |b|^2}{k}.
\ee
The pure (massless) ABJM model corresponds to the IR limit of M2-branes on $\mathbb R^{2,1}\times {\mathbb C}^4/{\mathbb Z}_k$ and
has as a gravity dual type IIA string theory on $AdS_4\times {\mathbb CP}^3$. In the massive case, the spacetime for M2-brane propagation is 
more complicated \cite{Mohammed:2010eb}, and the gravity dual even more so \cite{Auzzi:2009es,Mohammed:2010eb}, 
so we will not reproduce the formulas here.

To summarize, in this paper we have presented a consistent truncation of the ABJM model to a collective model of ${\cal O}(N)$ modes out of 
the ${\cal O}(N^2)$, reducing to an abelian Landau-Ginzburg model. We have also seen that we can map this process to a simple condensed matter 
model that reproduces the same general features. This provides a concrete step towards a well-defined AdS/CMT model, where there is a large $N$ theory for the condensed matter system, with a gravity dual, and yet the relevant physics is encoded in a simple abelian model. 

\section{Acknowledgements}
We thanks the anonymous referees for useful comments on an earlier version of this article that led to several clarifiying remarks. The work of HN is supported in part by CNPq grant 301219/2010-9. JM acknowledges support from an NRF Incentive Funding for Rated Researchers grant.

\end{document}